# Overdriven dislocation-precipitate interactions at elevated temperatures in aluminum

Amirreza Keyhani[*]

*The George W. Woodruff School of Mechanical Engineering, Georgia Institute of Technology, Atlanta, GA 30332-0405, USA*

January 25, 2018

**Abstract**

The two-dimensional dislocation dynamics approach has been recently used for analyzing plastic deformation in metals and alloys at elevated temperatures. The two-dimensional approach, however, only accounts for the dislocation climbing process, and it assumes that dislocation bypassing and shearing of precipitates are negligible. To examine the validity of this assumption, this study quantifies dislocation bypassing and shearing of precipitates in terms of critical resolved shear stress, interaction time, and thermal activation energy for various precipitate strength levels, temperatures, and precipitate spacings. This study uses a modified dislocation dynamics approach that accounts for shearable and non-shearable precipitates. Simulations focus on the overdriven dislocation dynamics regime wherein the climbing process is limited by fast interactions between dislocations and precipitates. The results show that even though the resolved shear stress level required for a dislocation to overcome an array of precipitates decreases at higher temperatures, the interaction time between the dislocation and the precipitates increases. In addition, the maximum ratio of thermal activation energy to the precipitate energy barrier is only 0.15.

**1. Introduction**

The subject of dislocation-obstacle interactions occupies a central place in the analysis of plastic deformation in metals and alloys. The development of computational methods for analyzing the movements and interactions of dislocations has paved the way for the statistics of dislocation-precipitate interactions. Early simulation methods were geometrical, and they modeled precipitates as dimensionless obstacles against dislocation movements. The need for more realistic models motivated the development of the dislocation dynamics approach (DD) [1-4]. This approach has been widely used for analyzing plasticity at the micron scale. However, the simulation of dislocation-precipitate interactions in DD has been a challenging problem.

---

[*]corresponding author
Email: akeyhani3@gatech.edu

Many studies devoted to dislocation-precipitate interactions are limited to the stress fields caused by precipitates. While some studies [5-8] introduced precipitates as spherical stress fields, other studies [9-11] evaluated the stress field resulting from matrix and precipitate shear modulus mismatch by applying the superposition principle, which decomposes the dislocation-precipitate interaction into two problems: a dislocation problem in an infinite homogeneous body and a correction problem representing the elastic stress field of the precipitates. Hence, the latter requires an extra numerical method such as the finite element method (FEM) or the boundary element method (BEM). The coupling of DD and an extra computational method complicates the solution of large systems with a random distribution of precipitates. The complications and disadvantages of modeling precipitates with the stress fields motivated Keyhani et al. [12-19] to develop a more efficient and robust methodology for modeling precipitates within the dislocation dynamics approach. In this methodology, if dislocation nodes encounter a precipitate, the nodes are locked until the shear stress level acting on the dislocation nodes exceeds the precipitate strength.

Dislocation-precipitate interaction scenarios depend on levels of temperature and resolved shear stress. At low temperatures, when a dislocation encounters a precipitate, it bends, so the related shear stress that the dislocation exerts on a precipitate increases. If this stress reaches a critical value, the dislocation overcomes precipitates via two mechanisms: passing by or passing through the precipitate. The first is when a dislocation forms a loop and passes by the highly resistant (impenetrable) precipitate. At high temperatures, an edge dislocation also can move out of its slip plane, known as the dislocation climbing process [20-23]. However, at high-stress levels relative to the critical resolved shear stress (CRSS), dislocation bypassing and shearing of precipitates dominate the thermally-assisted climbing mechanism. For further details about the temperature dependence of dislocation-obstacle interactions, see Refs. [24, 25].

The fact that the plastic response of metals and alloys is temperature dependent has spawned a large number of studies using computational [26-28] and experimental [29, 30] approaches to exploring the physics of plastic deformation at elevated temperatures. In the recent literature, several studies used the two-dimensional dislocation dynamics approach to analyze plastic deformation at elevated temperatures by incorporating the climbing process and adjusting elastic parameters and drag coefficients [31-33]. However, these studies neglected dislocation bypassing and shearing of precipitates. To provide quantitative information about dislocation bypassing and shearing of precipitates at elevated temperatures, the present study applies the author's recently developed computational framework [14] to quantify dislocation-precipitate interactions in the overdriven regime when the climbing process is limited because of fast interactions between dislocations and precipitates.

## 2. Modeling Approach

This study uses a recently proposed computational method [14] to model precipitates in the three-dimensional dislocation dynamics (DD) simulation code, DDLab [34]. Here, the dislocation dynamics approach and the methodology used for modeling precipitates are briefly reviewed. In the dislocation dynamics approach, a dislocation curve is discretized into straight-line segments, and each segment is defined by its two end nodes. The nodal velocities are calculated by solving mobility equations, and the dislocations' movements are performed by topological considerations. More details on DD can be found in [34]. In this study, simulations are based on elastic isotropy and constant drag coefficients.

A dislocation rounds a precipitate with a diameter larger than the precipitate's $(D' > D)$. As a result, the maximum stress that a dislocation can exert on a precipitate is

$$\tau_{\max} = \frac{2\mu b}{D'}, \quad (1)$$

where $\mu$ is the shear modulus of the matrix and $b$ is the magnitude of the Burgers vector. $D'$ is the diameter of the first Orowan loop [14],

$$D' = L + D - 2\pi L \beta \left[ \ln\left(\bar{D}/r_0\right) \right]^{-1}, \qu(2)$$

where $L$ is the internal distance between the two precipitates, $D$ is the precipitate diameter considered to be $100\,\text{nm}$ for all simulations, $r_0$ is the core radius of dislocation, and $\bar{D}=\left(D^{-1}+L^{-1}\right)^{-1}$. The constant $\beta$ depends on the Poisson's ratio, the dislocation core radius, and the dislocation line properties, which is equal to 0.75 [35, 36].

The precipitate resistance ratio $(R)$ is

$$R = \frac{\tau_p}{\tau_{max}}, \qquad (3)$$

where $\tau_p$ is the precipitate's strength against shearing. The resistance ratio is equal to unity for non-shearable precipitates, representing the Orowan mechanism, and less than unity for shearable precipitates. In the approach used here, when the distance of a node from the center of a precipitate is less than the radius of the first Orowan loop $(D'/2)$, we lock the node, and at each step, we compare the precipitate strength to the local shear stress exerted by the dislocation, the latter of which is related to the local curvature of the dislocation. If the local shear stress exerted by the dislocation exceeds the precipitate resistance, we release the node. This study does not account for the precipitates' evolution at high temperatures, and the precipitate resistance level is constant for each simulation.

The dislocation-precipitate interaction time is the interval that begins when a dislocation line initially starts to deform under the influence of precipitates and that ends when the dislocation line is about to pass by or through the precipitate array. Since the interaction time goes to infinity if the applied shear stress is equal to the critical resolved shear stress $(\tau_{CRSS})$, we introduce $\tau_c = 1.05\,\tau_{CRSS}$, and $t_c$ is the corresponding time at the critical state for $\tau_c$. Figure 1 illustrates the simulation scheme. We perform simulations for Al, an FCC crystal, and four precipitate spacing ratios $(L/D)$ of 2.5, 5, 7.5, and 10. The level of applied shear stress and the temperature are constant in each simulation. For each level of temperature, elastic constants and drag coefficients are modified based on the values presented in Fig. 2 [31]. Since the Burgers vector only increases by 2% when the temperature increases from 0 K to 900 K [37], the simulations do not account for variations in the Burgers vector. In DD simulations, the maximum length of a dislocation segment and the maximum movement of a segment in each step are 10 nm.

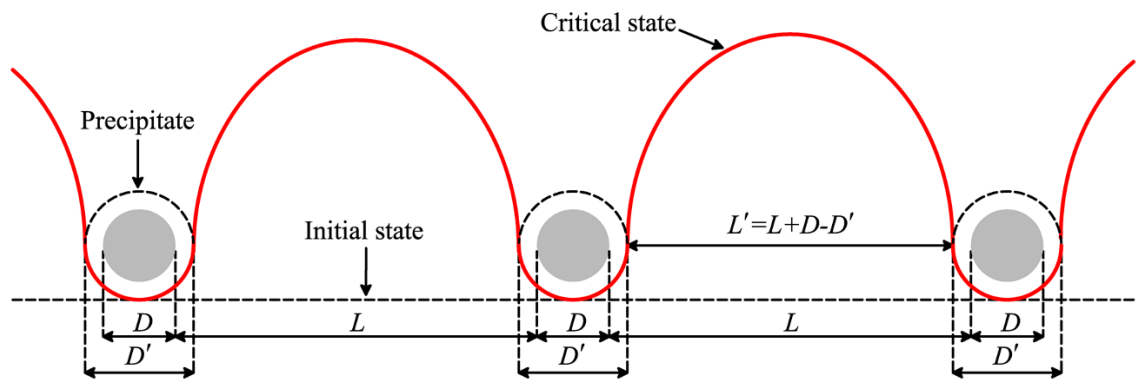

Fig. 1. Simulation scheme: a dislocation line encounters an array of identical, equally-spaced precipitates.

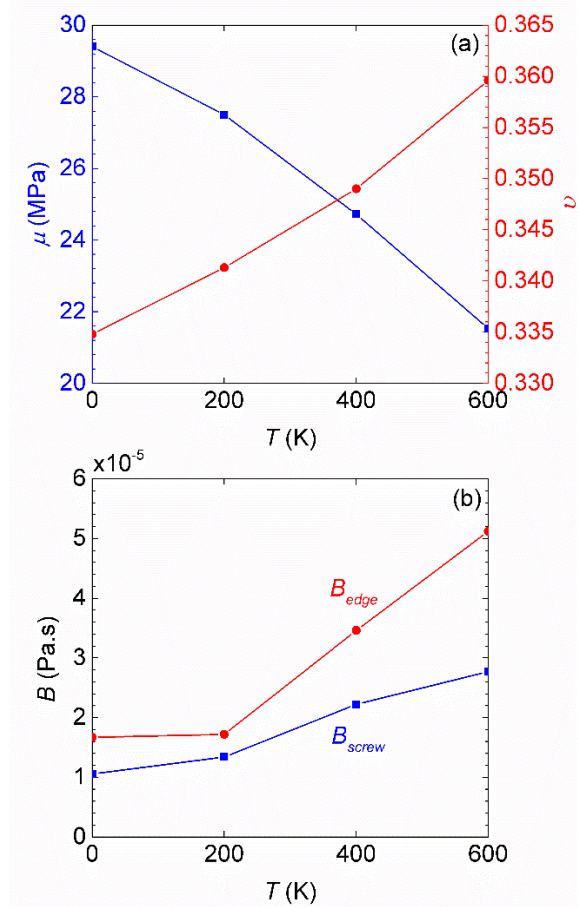

Fig. 2. Variations in macroscopic and microscopic properties of aluminum caused by temperature changes: (a) shear modulus $(\mu)$ and Poisson's ratio $(\upsilon)$, and (b) drag coefficients for edge and screw dislocations [31].

## 3. Results and Discussion

This study analyzes several aspects of short-range interactions between which is initially an edge dislocation and an array of equally-spaced identical precipitates over the temperature range of 0-600 K. These aspects include (1) variations in critical shear stress and the critical interaction time resulting from temperature changes, (2) evolutions of the interaction time at shear stress levels higher than critical shear stress, and (3) effects of temperature and the precipitate spacing ratio on the contribution of thermal activation energy.

*3.1. Critical resolved shear stress (CRSS) and the critical interaction time at elevated temperatures*

At absolute zero temperature, the barriers resisting the glide of a dislocation can be overcome only if the resolved shear stress exceeds the local peak glide resistance. At temperatures above absolute zero, however, thermal fluctuations redistribute energy resulting in shearing or bypassing of barriers at resolved shear stress levels below peak resistance levels. Figures 3(a)-3(c) illustrate critical shear stress (i.e., $\tau_c(T) = 1.05\tau_{CRSS}(T)$) for various precipitate spacings $(L/D)$, temperature levels $(T)$, and for precipitate resistance ratios of 1, 0.8, and 0.6, respectively. The results show that an increase in temperature causes a decrease in critical resolved shear stress. This decrease is more significant for smaller precipitate spacing ratios when the critical shear stress is higher. To pass either by or through an array of precipitates with a known resistance level, a dislocation line bows to a certain radius. The required shear stress level for bowing a dislocation line to radius $\rho$ has a direct relationship with the shear modulus $(\mu)$, $\tau = \beta\mu b/\rho$ [38]. At elevated temperatures, the shear modulus decreases; therefore, the corresponding required shear stress level for bowing a dislocation line to radius $\rho$ decreases.

Figures 3(d)-3(f) illustrate the evolution of the critical interaction time $(t_c)$ over the temperature range of 0-600 K for various precipitate spacings and resistance levels. These figures show that an increase in temperature causes the critical interaction time to increase. At constant stress, if temperature increases, the dislocation velocity decreases as a result of an increase in drag coefficients [39]. In addition, since the critical applied shear stress level decreases at the elevated temperatures, the applied force on the dislocation decreases. The decrease in the driving force and the velocity of a dislocation result in a significant increase in the interaction time. This increase in the interaction time is more notable for higher precipitate spacings.

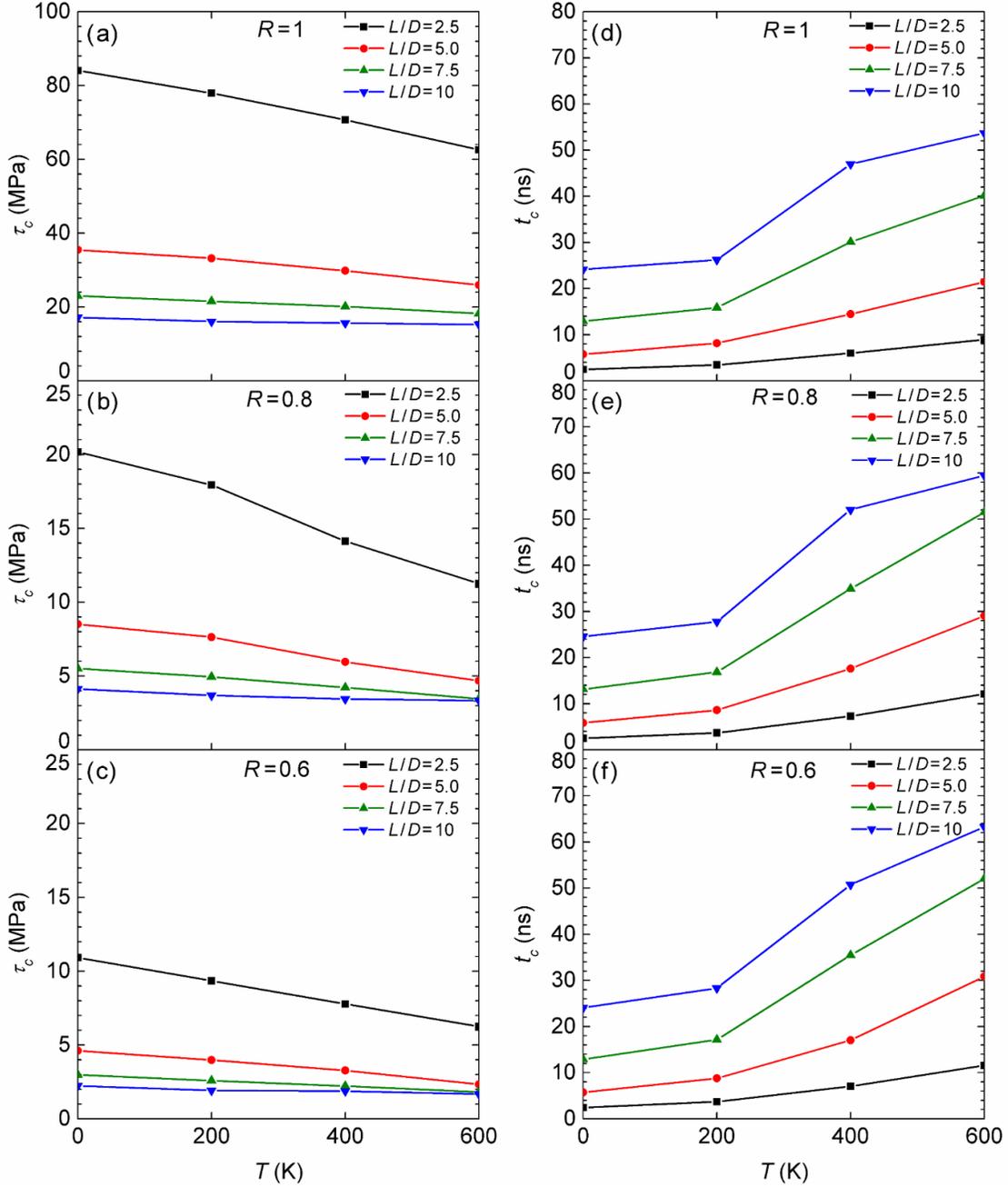

Fig. 3. Variations in critical resolved shear stress and the interaction time resulting from changes in temperature $(T)$, internal spacing ratio $(L/D)$, and the precipitate resistance ratio $(R)$: (a)-(c) critical resolved shear stress, and (d)-(f) the critical interaction time.

## 3.2. Evolutions of the interaction time with shear stress

The previous section quantifies the interaction time for cases when applied shear stress is equal to critical shear stress (i.e., $\tau_c = 1.05\tau_{CRSS}$), referred to as "critical interaction time." Figures 4-6, however, illustrate the interaction time $(t_i)$ when the applied resolved shear stress $(\tau_a)$ is higher than the critical resolved shear stress $(\tau_c)$. These figures present the interaction time for various applied shear stress levels, precipitate spacings, resistance ratios, and temperatures. Specifically, Fig. 4 shows the interaction time for non-shearable precipitates. As drag coefficients increase at elevated temperatures, for a known precipitate spacing ratio $(L/D)$, an increase in temperature causes the interaction time to increase. At applied resolved shear stress levels slightly higher than $\tau_c$, the interaction time sharply decreases. This decrease is more significant for higher precipitate spacing ratios. Figures 5 and 6 show the interaction time for shearable precipitates with resistance ratios of 0.8 and 0.6, respectively. The trend of the interaction time for shearable precipitates is similar to that for non-shearable precipitates; however, the decrease in the interaction time at $\tau_a > \tau_c$ is more significant for weaker precipitates.

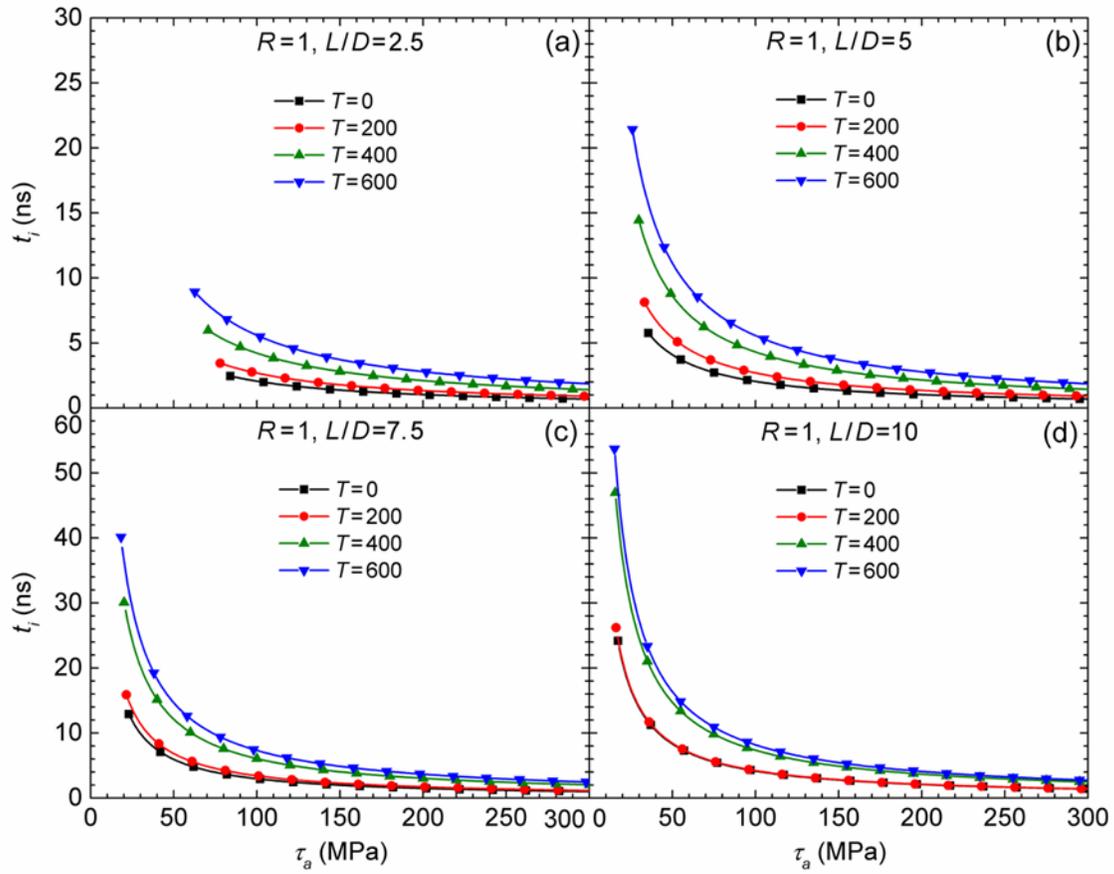

Fig. 4. Evolution of the interaction time versus applied resolved shear stress, and the precipitate spacing $(L/D)$ for a resistance ratio of $R=1$: (a) $L/D=2.5$, (b) $L/D=5$, (c) $L/D=7.5$, and (d) $L/D=10$.

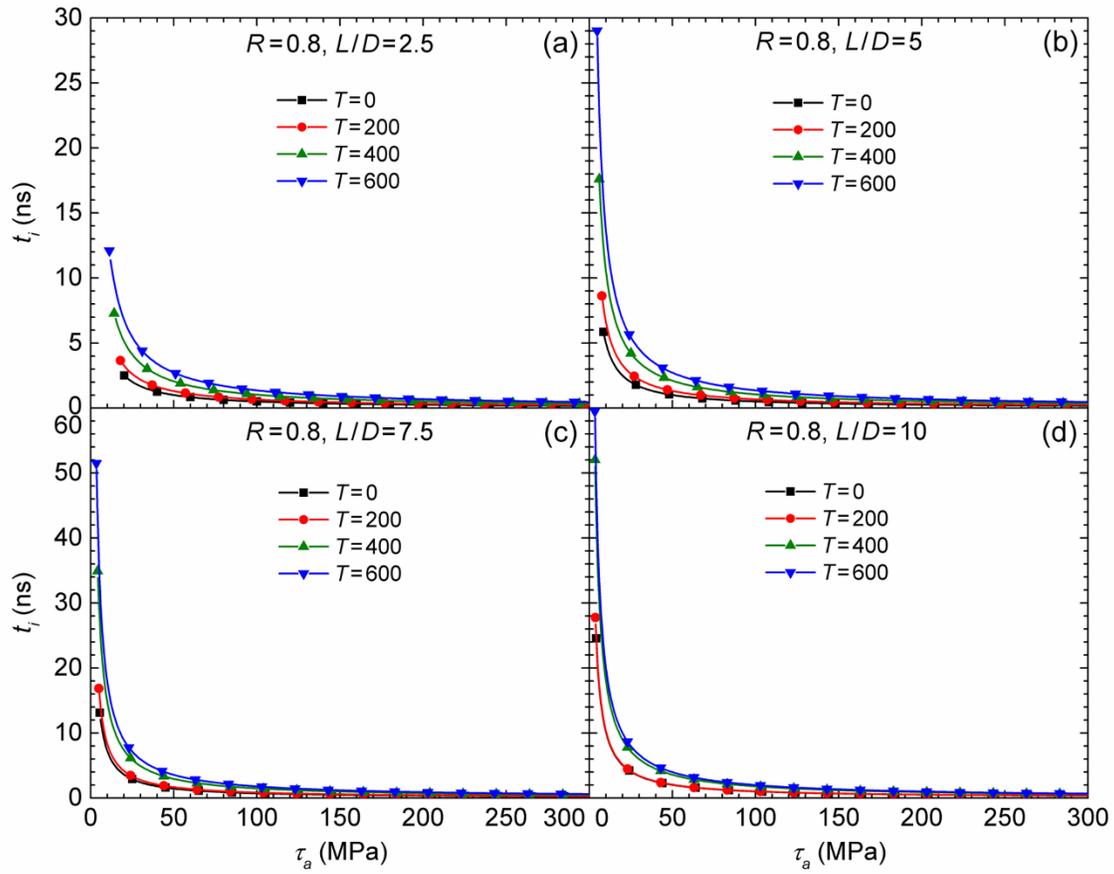

Fig. 5. Evolution of the interaction time versus applied resolved shear stress, and the precipitate spacing $(L/D)$ for a resistance ratio of $R = 0.8$: (a) $L/D = 2.5$, (b) $L/D = 5$, (c) $L/D = 7.5$, and (d) $L/D = 10$.

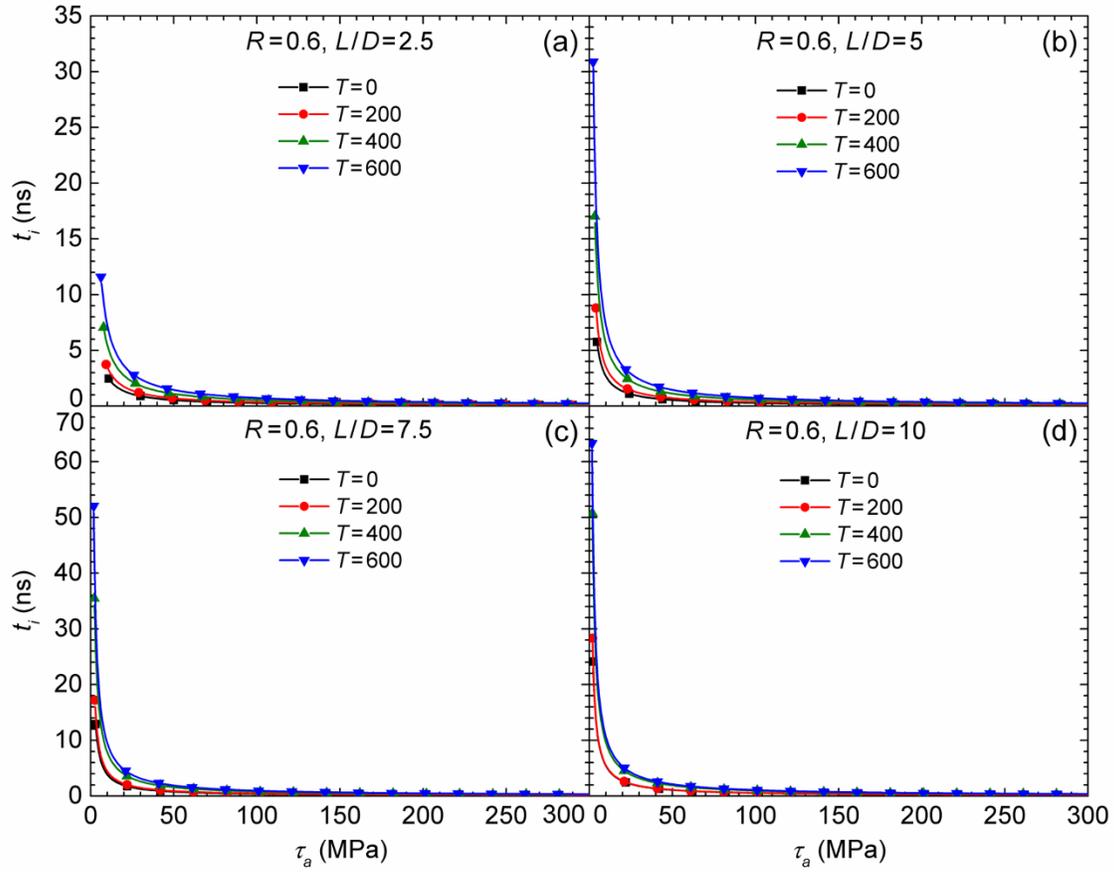

Fig. 6. Evolution of the interaction time versus applied resolved shear stress, and the precipitate spacing $(L/D)$ for a resistance ratio of $R = 0.6$: (a) $L/D = 2.5$, (b) $L/D = 5$, (c) $L/D = 7.5$, and (d) $L/D = 10$.

### 3.3. Contribution of thermal activation energy

The energy barrier against penetration of a dislocation through an array of precipitates $(\Delta F^*)$ can be overcome by thermal activation and applied stress,

$$\Delta F^* = \Delta G^* + \Delta W^*. \tag{4}$$

$\Delta G^*$ (i.e., activation free enthalpy, or Gibbs free energy of activation) is energy supplied by thermal fluctuations at constant temperature and stress. $\Delta W^*$ is the work done by stress applied during the interaction. The ratio of thermal activation energy to the energy barrier $\left(\Delta G^*/\Delta F^*\right)$ shows the contribution of thermal activation [39, 40],

$$\frac{\Delta G^*}{\Delta F^*} = \left[1-\left(\frac{\tau_c(T)}{\tau_c(0)}\right)^{\frac{1}{2}}\right]^{\frac{3}{2}}, \qquad (5)$$

where $\tau_c(T)$ is the resolved shear stress level required to overcome an array of precipitates at absolute temperature $T$. Figure 7 presents $\Delta G^*/\Delta F^*$ for various precipitate spacings, precipitate resistance levels, and temperatures. In general, an increase in the temperature causes thermal activation energy to increase. The results show that the contribution of thermal activation slightly increases up to a precipitate spacing ratio of 5 and then decreases when the precipitate spacing ratio increases. In addition, the contribution of thermal activation is higher for weaker precipitates, and the maximum ratio of thermal activation energy to the precipitate energy barrier is 0.15. Figure 8 summarizes the results of Figs. 7(a)-7(c). The filled areas in Fig. 8 show the variations in $\Delta G^*/\Delta F^*$ among cases with similar precipitate resistance levels but different precipitate spacings. These variations increase at higher temperatures.

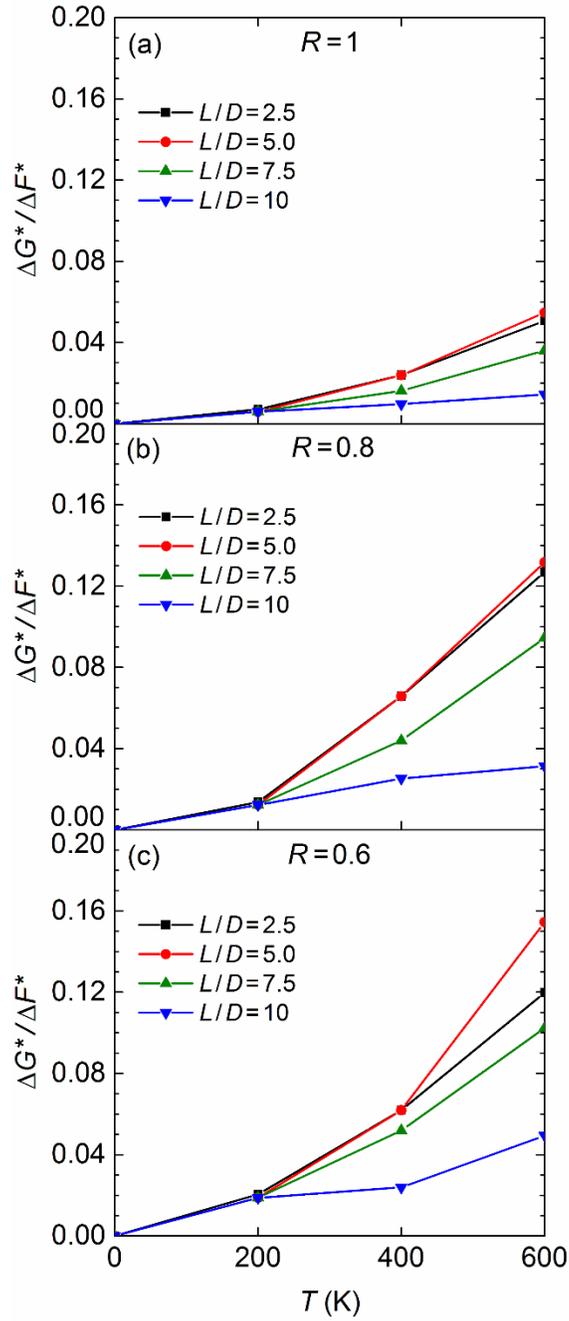

Fig. 7. Ratio of thermal activation energy to the precipitate energy barrier versus temperature for various precipitate spacings $(L/D)$ and precipitate resistance, levels including (a) $R=1$, (b) $R=0.8$, and (c) $R=0.6$.

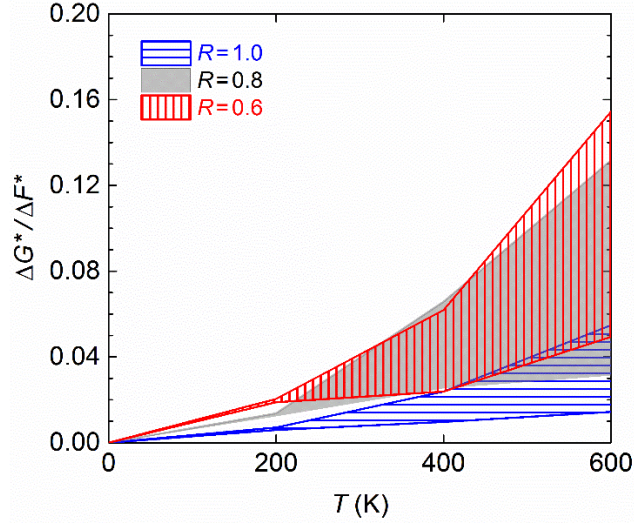

Fig. 8. Ratio of thermal activation energy to the precipitate energy barrier. The filled areas show the variations in $\Delta G^*/\Delta F^*$ among cases with a similar precipitate resistance level but different precipitate spacings.

## 4. Conclusion

This study examined the interaction of an edge dislocation line with an array of collinear equally-spaced precipitates at elevated temperatures using a modified line dislocation dynamics approach. In the simulations, the macroscopic and microscopic properties of the material were updated based on the temperature level. This study quantified the dislocation-precipitate interactions in terms of critical resolved shear stress, interaction time, and thermal activation energy for various temperatures, precipitate spacings, and precipitate resistance levels. The results reveal an opposite effect of temperature on critical shear stress and the dislocation-precipitate interaction time. At higher temperatures, the shear stress level required for a dislocation to overcome an array of precipitates decreases, whereas the interaction time increases. The results show that at applied shear stress levels slightly higher than the critical shear stress, the interaction time sharply decreases. This decrease is more significant for larger precipitate spacings or lower precipitate resistance levels. In addition, the ratio of thermal activation energy to the precipitate energy barrier decreases when the precipitate resistance level or the spacing increases. Finally, despite significant variations in critical resolved shear stress and the critical interaction time at elevated temperatures, the maximum ratio of thermal activation energy to the precipitate energy barrier is only 0.15 in the overdriven regime.